\documentclass{kapedbk}

\usepackage{graphicx}

\draft

\let\footnote\savefootnote

\normallatexbib

\begin{document}

\articletitle[Quantum Physics in Inertial and Gravitational Fields ]{Quantum Physics in
Inertial
\\ and Gravitational Fields}

\chaptitlerunninghead{Quantum Physics}

\author{G. Papini}
\affil{Department of Physics, University of Regina, Regina, Sask. S4S 0A2, Canada.\\
International Institute for Advanced Scientific Studies, 84019 Vietri sul Mare (SA),
Italy.} \email{papini@uregina.ca}

\begin{abstract}
Covariant generalizations of well-known wave equations predict the existence of
inertial-gravitational effects for a variety of quantum systems that range from
Bose-Einstein condensates to particles in accelerators. Additional effects arise in
models that incorporate Born reciprocity principle and the notion of a maximal
acceleration. Some specific examples are discussed in detail.

\end{abstract}



\section{Introduction}
\setcounter{equation}{0}
 The interaction of quantum systems with external inertial and
gravitational fields is of interest in studies regarding the ultimate structure of
space-time. Covariant generalizations of well known wave equations provide examples of
effects involving classes of quantum systems in conditions remote from the onset of
quantum gravity, hence amenable, it is hoped, to observation. For this purpose,
Schroedinger, Klein-Gordon, Maxwell-Proca and Dirac equations have been frequently
discussed in the literature. The Landau-Ginzburg and Gross-Pitaevskii equations should
also be added to this group because of the peculiar properties of charged and neutral
Bose-Einstein condensates. As shown in Section 2, these equations can be solved exactly
to first order in the weak field approximation (WFA), if the solutions of the
corresponding field free equations are known. The same procedure can also be applied to
de Rham, Maxwell-Proca and Dirac equations.

The interaction of quantum systems with external inertial and gravitational fields
produces quantum phases. Though these are in general path-dependent, phase differences
are observable, in principle, by means of interferometers. Section 2 refers to this first
group of effects. An explicit calculation of the phase difference due to the
Lense-Thirring (LT) effect is added for pedagogical reasons.

A second group of effects, considered in Sections 3, is derived from effective
Hamiltonians for the motion of fermions in accelerators and storage rings. It deals
essentially with spin-rotation coupling, its non-universal character and its invariance
under parity and time reversal.

The problems considered in Section 4 stem from attempts to incorporate Born reciprocity
theorem into the structure of space-time. They are related to the notion of a maximal
acceleration (MA), whose presence, frequently discussed in both classical and quantum
contexts and in string theory, plays the role of a field regulator while preserving the
continuous structure of space-time. The MA corrections to the Lamb shift of one-electron
atoms and ions, also discussed in Section 4, are comparable in magnitude with those of
quantum electrodynamics of order seven in the fine structure constant and are
not,therefore, negligible. Section 5 contains a summary.

\section{Quantum phases}

\subsection{Landau-Ginzburg and Gross-Pitaevskii equations}

\setcounter{equation}{0}

In view of the wide variety of interferometers presently in use or under development, it
is convenient to study systems whose wave functions satisfy the equation
\begin{eqnarray}\label{2a}
\left[ (\nabla_{\mu} + i \frac{e}{c} A_{\mu})^{2} + \frac{m^{2}c^{2}}{\hbar^{2}}
\right]\Phi (x) = \beta \mid \Phi (x) \mid ^{2} \Phi (x),
\end{eqnarray}
where $ \nabla_{\mu}$ indicates covariant differentiation, $\beta$ is a constant and
$A_{\mu} (x)$ represents the total electromagnetic potential of all external and gravity
induced fields present. Eq.(\ref{2a}) is the fully covariant version of the
Landau-Ginzburg equation \cite{cai1}. It reduces to the Gross-Pitaevskii equation when
$A_{\mu}$ vanishes and to the Klein-Gordon equation when $\beta = 0$.  It is therefore
well suited to discuss a number of systems, from superfluids \cite{pap1} and
Bose-Einstein condensates \cite{chiao}, to scalar particles.  If, in particular, heavy
fermion systems admit minimal coupling \cite{Varma}, then Eq.(\ref{2a}) may be used in
this case too with the added advantage of a much larger effective coupling in mixed
gravity-electromagnetism interaction terms.

In the WFA $ g_{\mu\nu}=\eta_{\mu\nu}+\gamma_{\mu\nu}$, where $ \gamma_{\mu\nu}$ is the
metric deviation, $ |\gamma_{\mu\nu}|\ll 1$ and the signature of $ \eta_{\mu\nu}$ is $
-1$. To first order, Eq.(\ref{2a}) becomes ( $\hbar=c=G=1$)
\begin{equation}\label{2b}
[(\eta^{\mu\nu}-\gamma^{\mu\nu})\partial_{\mu}\partial_{\nu}-(\gamma^{\alpha\mu}-1/2
\gamma_{\sigma}^{\sigma}\eta^{\alpha\mu}),_{\mu}\partial_{\alpha}+ m^{2}- \beta \mid
\Phi\mid ^{2} ]\Phi{(x)}=0 {.}
\end{equation}
It is useful to start with the ansatz
\begin{equation}\label{2c}
\Phi(x)=exp \left(-i\chi\right) \phi_{0}(x)\simeq (1-i\chi) \phi_{0}(x) ,
\end{equation}
where $\phi_{0}(x)$ is a field quantity to be determined below
 and
\begin{eqnarray}\label{2.4}
i \chi \phi_{0}&=& \frac{1}{4}\int_{P}^{x}dz^{\lambda}
(\gamma_{\alpha\lambda,\beta}(z)-\gamma_{\beta\lambda,\alpha}(z))[(x^{\alpha}-z^{\alpha})
\partial^{\beta}- \nonumber \\ &&(x^{\beta}-z^{\beta})\partial^{\alpha}]-
\frac{1}{2}\int_{P}^{x}dz^{\lambda}\gamma_{\alpha\lambda}(z)\partial^{\alpha}]\phi_{0}.
\end{eqnarray}
Because coordinates play the role of parameters in relativity, phase (2.4) is sometimes
referred to as the gravitational Berry phase \cite{cai2}.

It is easy to prove by differentiation that (2.4) leads to
\begin{eqnarray}\label{2e}
i\partial_{\mu}(\chi\phi_{0})&=&\frac{1}{4}\int^{x}_{P}dz^{\lambda}
(\gamma_{\alpha\lambda,\beta}(z)-
\gamma_{\beta\lambda,\alpha}(z))[\delta_{\mu}^{\alpha}\partial^{\beta}-\delta_{\mu}^{\beta}
\partial^{\alpha}]\phi_{0}(x)+\nonumber\\
& & \frac{1}{4}\int_{P}^{x}dz^{\lambda}(\gamma_{\alpha\lambda,\beta}(z)-
\gamma_{\beta\lambda,\alpha}(z)) [(x^{\alpha}-z^{\alpha})\partial^{\beta}-\nonumber\\ & &
(x^{\beta}-z^{\beta})\partial^{\alpha}]
\partial_{\mu}\phi_{0}(x)-
\frac{1}{2}\int_{P}^{x}dz^{\lambda}
\gamma_{\alpha\lambda}(z)\partial^{\alpha}\partial_{\mu}\phi_{0}(x)-\nonumber\\ & &
\frac{1}{2} \gamma_{\alpha\mu}(x)
\partial^{\alpha}\phi_{0}(x) ,
\end{eqnarray}
from which one gets
\begin{eqnarray}\label{2.6}
i\partial_{\mu}\partial^{\mu}(\chi\phi_{0})&=& -im^2 \chi\phi_{0} + i \chi( \beta \mid
\phi_{0}\mid^{2}\phi_{0})- \nonumber\\ & &
\gamma_{\mu\alpha}\partial^{\mu}\partial^{\alpha}\phi_{0}
 - (\gamma^{\beta\mu}-
\frac{1}{2}\gamma_{\sigma}^{\sigma}\eta^{\beta\mu}),_{\mu}\partial_{\beta}\phi_{0}.
\end{eqnarray}
By substituting (2.6) and (\ref{2c}) into (\ref{2b}) one finds, to lowest order,
\begin{eqnarray}\label{2g}
[(\eta^{\mu\nu}- \gamma^{\mu\nu})\partial_{\mu}\partial_{\nu}&+& m^2 - \beta \mid
\Phi\mid ^{2} ]\Phi{(x)}=
  \nonumber \\ & & [\eta^{\mu\nu}\partial_{\mu}\partial_{\nu}+
m^{2}- \beta \mid \phi_{0}\mid ^{2} ]\phi_{0}{(x)}+ \nonumber \\ & & \beta \left[\mid
\phi_{0}\mid^{2}\left(i \chi\phi_{0}\right)
 - i \chi\left( \mid
\phi_{0}\mid^{2}\phi_{0}\right)\right] ,
\end{eqnarray}
where use has been made of the Lanczos-DeDonder gauge condition
\begin{equation}\label{2g*}
\gamma_{\alpha, \nu}^{\nu}-\frac{1}{2}\gamma^{\sigma}_{\sigma, \alpha}=0 .
\end{equation}
 Eq.(\ref{2c}) therefore is a solution of (\ref{2b}) exact to first order if
\begin{eqnarray}\label{2h}
[\eta^{\mu\nu}\partial_{\mu}\partial_{\nu}+ m^{2}- \beta \mid \phi_{0}\mid ^{2}
]\phi_{0}{(x)} \mid \phi_{0}\mid^{2}+ \nonumber \\ \beta\left[\left(i \chi\phi_{0}
\right)- i \chi\left(\mid \phi_{0}\mid^{2}\phi_{0}\right)\right]= 0.
\end{eqnarray}
In problems where $ \mid \phi_{0}\mid^{2}$ is constant, $ \phi_{0}$ satisfies the
Ginzburg-Landau equation
\begin{equation}\label{2i}
[\eta^{\mu\nu}\partial_{\mu}\partial_{\nu}+ m^{2}- \beta \mid \phi_{0}\mid ^{2}
]\phi_{0}{(x)}=0.
\end{equation}
When $ \beta = 0$,  (\ref{2a}) becomes the covariant Klein-Gordon equation and (\ref{2i})
the Klein-Gordon equation in Minkowski space.

 For a
closed path in space-time one finds \cite{cai1}
\begin{equation}\label{2l}
i\Delta\chi\phi=\frac{1}{4}\int_{\Sigma_{p}}{R_{\mu\nu\alpha\beta}L^{\alpha\beta}
d\tau^{\mu\nu}}\phi_{0} {,}
\end{equation}
where $ \Sigma_{p}$ is the surface bound by the closed path, $L^{\alpha\beta}$is the
angular momentum of the particle of mass $m$, and $R_{\mu\nu\alpha\beta}$ is the
linearized Riemann tensor
\begin{equation}\label{2m}
R_{\mu\nu\alpha\beta}=\frac{1}{2}\left(\gamma_{\mu\beta,\nu\alpha}+\gamma_{\nu\alpha,\mu\beta}-
\gamma_{\mu\alpha,\nu\beta}-\gamma_{\nu\beta,\mu\alpha}\right).
\end{equation}

 Result (\ref{2l}) is manifestly gauge invariant.
 The effect of the
electromagnetic field can also be incorporated in the phase factor in a straight-forward
way by adding to $ i\chi$ the term $ ie \int^{x}_{P} dz^{\lambda} A_{\lambda}(z)$. The
additional phase difference is $ e \int_{\Sigma_{p}} F_{\mu \nu} d \tau^{\mu \nu}$ where
$F_{\mu \nu} = -A_{\mu, \nu} + A_{\nu, \mu}$.

\subsection{de Rahm and Maxwell equations}
The de Rahm wave equation
\begin{equation}\label{2n}
\nabla_{\nu}\nabla^{\nu}A_{\mu}-R_{\mu\sigma}A^{\sigma}=0 {,}
\end{equation}
where $\nabla_{\mu}A^{\mu}=0$, becomes, in the WFA and in the gauge (\ref{2g*}),
\begin{eqnarray} \label{2o*}
\nabla_{\nu}\nabla^{\nu}A_{\mu}-R_{\mu\sigma}A^{\sigma}&\simeq&(\eta^{\sigma\alpha}-\gamma^{\sigma\alpha})
A_{\mu,\alpha\sigma}-\nonumber \\ &&(\gamma_{\sigma\mu,\nu}+
\gamma_{\sigma\nu,\mu}-\gamma_{\mu\nu,\sigma})A^{\sigma,\nu}=0 {.}
\end{eqnarray}
This equation has the solution
\begin{equation}\label{2p}
A_{\mu}= exp(-i\xi)\simeq (1-i\xi)a_{\mu},
\end{equation}
where
\begin{eqnarray}\label{2q}
i\xi a_{\mu}(x)&=&\frac{1}{4}\int_{P}^{x}dz^{\lambda}(\gamma_{\alpha\lambda,\beta}(z)-
\gamma_{\beta\lambda,\alpha}(z))[(x^{\alpha}-z^{\alpha})\partial^{\beta}a_{\mu}(x)
-\nonumber \\ && (x^{\beta}-z^{\beta})\partial^{\alpha}a_{\mu}(x)]+
 \frac{1}{2}\int_{P}^{x}dz^{\lambda}(\gamma_{\mu\lambda,\sigma}(z)-
\gamma_{\sigma\lambda,\mu}(z))a^{\sigma}-\nonumber \\
&&\frac{1}{2}\int_{P}^{x}dz^{\lambda} \gamma_{\alpha\lambda}(z)
\partial^{\alpha}a_{\mu}(x)-
\int_{P}^{x}dz^{\lambda}\gamma_{\alpha\mu}(z)\partial_{\lambda}a^{\alpha}(x) ,
\end{eqnarray}
$\partial_{\nu}\partial^{\nu}a_{\mu}=0$ and $\partial^{\nu}a_{\nu}=0$. If $R_{\mu\sigma}$
is negligible, then Eq.(\ref{2n}) becomes Maxwell wave equation and the phase operator
$\xi$ can also be written in the form \cite{pap1}
\begin{eqnarray}\label{2r}
 \xi&=&\frac{1}{4}\int_{P}^{x}dz^{\lambda}(\gamma_{\alpha\lambda,\beta}(z)-
\gamma_{\beta\lambda,\alpha}(z))J^{\alpha\beta}-\nonumber\ \\ & &
\frac{1}{2}\int_{P}^{x}dz^{\lambda}\gamma_{\alpha\lambda}(z)\partial^{\alpha}-
\frac{1}{2}\int_{P}^{x}dz^{\lambda}\gamma_{\alpha\beta,\lambda}(z)T^{\alpha\beta}{,}
\end{eqnarray}
where $J^{\alpha\beta}=L^{\alpha\beta}+S^{\alpha\beta}$ is the total angular momentum,
$(S^{\alpha\beta})^{\mu\nu}=-i(g^{\mu\alpha}g^{\nu\beta}-g^{\mu\beta}g^{\nu\alpha})$ is
the spin-1 operator and $(T^{\alpha\beta})^{\mu\nu}\equiv
-i\frac{1}{2}(g^{\mu\alpha}g^{\nu\beta}+g^{\mu\beta}g^{\nu\alpha})$. All spin effects are
therefore contained in the $S^{\alpha\beta}$ and $T^{\alpha\beta}$ terms. For a closed
path one can again find a gauge invariant equation similar to (\ref{2l}).

The procedure discussed can be easily extended to massive vector particles.

\subsection{Covariant Dirac equation}
Some of the most precise experiments in physics involve spin-1/2 particles. They are very
versatile tools that can be used in a variety of experimental situations and energy
ranges while still retaining a non-classical behaviour. Within the context of general
relativity, De Oliveira and Tiomno \cite{de} and Peres \cite{peres} conducted
comprehensive studies of the fully covariant Dirac equation which takes the form
\begin{equation}\label{cde}
[i\gamma^{\mu}(x)D_{\mu}-m]\Psi(x)=0 {,}
\end{equation}
where $D_{\mu}=\nabla_{\mu}+i\Gamma_{\mu}$. The generalized matrices $ \gamma^{\mu}(x)$
satisfy the relations $\{\gamma^{\mu}(x),\gamma^{\nu}(x)\}=2 g^{\mu\nu}(x)$,
$D_{\mu}\gamma_{\nu}(x)=
\nabla_{\mu}\gamma_{\nu}(x)+i[\Gamma_{\mu}(x),\gamma_{\nu}(x)]=0$  and are related to the
usual Dirac matrices $ \gamma^{\hat{\alpha}}$ by means of the vierbeins $
e^{\mu}_{\hat{\alpha}}(x)$. The spin connection $\Gamma^{\mu}$ is
\begin{equation}\label{sin1}
\Gamma_\mu  \ = \ \frac{i}{4} \gamma^\nu (\nabla_\mu \gamma_\nu) \ = \ - \frac{1}{4}
\sigma^{\hat{\alpha} \hat{\beta}} e^\nu{}_{\hat{\alpha}}
 (\nabla_ \mu e_{\nu \hat{\beta}}),
\end{equation}
where $ \sigma^{\hat{\alpha} \hat{\beta}} = \frac{i}{2} [ \gamma^{\hat{\alpha}},
\gamma^{\hat{\beta}} ]$ . Particularly interesting is the case of acceleration and
rotation \cite{hehl}\cite{singh}. In this instance it is possible to define a local
co-ordinate frame according to an orthonormal tetrad with three-acceleration $\vec{a}$
along a particle's world-line and three-rotation $\vec{\omega}$ of the spatial triad,
subject to Fermi-Walker transport. This tetrad $\vec{e}_{\hat{\mu}}$, is related to the
general co-ordinate tetrad $\vec{e}_\mu$ by
\begin{equation}\label{sin2}
\vec{e}_{\hat{0}}  =   \left(1 + \vec{a} \cdot \vec{x} \right)^{-1} \left[ \vec{e}_{0} -
 (\vec{\omega} \times \vec{x})^k \vec{e}_k \right],
\vec{e}_{\hat{\imath}}  =  \vec{e}_i . \end{equation}
The corresponding vierbeins relating the two frames are then
\begin{eqnarray}\label{sin3}
e^0{}_{\hat{0}} & = &  \left(1 + \vec{a} \cdot \vec{x} \right)^{-1}, e^k{}_{\hat{0}}  = -
\left(1 + \vec{a} \cdot \vec{x} \right)^{-1} \epsilon^{ijk} \, \omega_i \, x_j, \nonumber
\\ e^0{}_{\hat{\imath}} &=& 0, e^k{}_{\hat{\imath}} = \delta^k{}_i .
\end{eqnarray}
Similarly, by inverting (\ref{sin2}), we find the inverse vierbeins
\begin{eqnarray}\label{sin4}
e^{\hat{0}}{}_{0} & = &  \left(1 + \vec{a} \cdot \vec{x} \right), e^{\hat{k}}{}_{0} =
 \epsilon^{ijk} \, \omega_i \, x_j, \nonumber \\ e^{\hat{0}}{}_{i} & = & 0 {,}
\  e^{\hat{k}}{}_{i}  =  \delta^k{}_i.
\end{eqnarray}
The vierbeins satisfy the orthonormality conditions
\begin{equation}\label{sin5}
\delta^{\hat{\alpha}}{}_{\hat{\mu}}  =   e^\nu{}_{\hat{\mu}} e^{\hat{\alpha}}{}_{\nu},
 \delta^\alpha{}_{\mu}  =   e^{\hat{\nu}}{}_{\mu}
e^\alpha{}_{\hat{\nu}}. \end{equation}
It follows that the metric tensor components are
\begin{eqnarray}\label{sin6}
g_{00} & = & \left(1 + \vec{a} \cdot \vec{x} \right)^2 + \left[ \left(\vec{\omega} \cdot
\vec{\omega} \right) \left(\vec{x} \cdot \vec{x} \right) - \left(\vec{\omega} \cdot
\vec{x} \right)^2 \right],\nonumber \\ g_{0j} & = & - \left(\vec{\omega} \times \vec{x}
\right)_j  {,} \ g_{jk}  =  \eta_{jk}.
\end{eqnarray} One also finds
\begin{equation}\label{sin7}
\Gamma_0  =  - \frac{i}{2} ( \vec{a} \cdot \vec{\alpha} ) -  \vec{\omega} \cdot
\vec{\sigma}  {,} \ \Gamma_j  =  0. \end{equation}
 By
using the definitions $ \Psi(x)=S \tilde{\Psi}(x) $, $
S=exp(-i\int_{P}^{x}dz^{\lambda}\Gamma_{\lambda}(z)) $ and $
\tilde{\gamma}^{\mu}(x)=S^{-1}\gamma^{\mu}(x)S$, in (\ref{cde}) one finds \cite{singh}
\begin{equation}\label{cdt}
 [i\tilde{\gamma}^{\mu}(x)\nabla_{\mu}-m]\tilde{\Psi}=0 {.}
 \end{equation}
By substituting $ \tilde{\Psi}=[-i\tilde{\gamma}^{\alpha}(x)\nabla_{\alpha}-m]\psi'$ into
(\ref{cdt}) , one obtains
\begin{equation}\label{ckg}
(g^{\mu\nu}\nabla_{\mu}\nabla_{\nu}+m^2)\psi'=0
 \end{equation}
 which, as shown above, has
the WFA solution $\psi'=exp(-i\chi)\psi_{0}$, where $\psi_{0}$ is a solution of the
Klein-Gordon equation in Minkowski space. It is again possible to show that for a closed
path the total phase difference experienced by the Dirac wave function is gauge invariant
and is given by $ \frac{1}{4}\int R_{\mu\nu\alpha\beta}J^{\alpha\beta}d\tau^{\mu\nu} $,
where the total angular momentum is now $ J^{\alpha\beta}=L^{\alpha\beta}+
\sigma^{\alpha\beta} $, $
\sigma^{\alpha\beta}=-\frac{1}{2}[\gamma^{\alpha},\gamma^{\beta}] $ and $\gamma^{\beta}$
represents a usual, constant Dirac matrix \cite{caipap2}. It then follows that the Dirac
Hamiltonian in the general co-ordinate frame is, to first-order in $\vec{a}$ and
$\vec{\omega}$,
\begin{equation}\label{sin8}
H \ \approx \  ( \vec{\alpha} \cdot \vec{p} ) + m \beta + V(\vec{x}),
\end{equation}
where
\begin{eqnarray}\label{sin9}
V(\vec{x}) & = & \frac{1}{2}[( \vec{a} \cdot \vec{x} ) ( \vec{\alpha} \cdot \vec{p} ) + (
\vec{\alpha} \cdot \vec{p} )( \vec{a} \cdot \vec{x} )]+ m(\vec{a} \cdot \vec{x}) \beta -
\nonumber \\ &   & \vec{\omega} \cdot (\vec{L} + \vec{S})  +  \vec{\alpha} \cdot
(\vec{\nabla} \Phi_{\rm G}) +  \nabla_0 \Phi_{\rm G},
\end{eqnarray}
the $\vec{\alpha}, \beta, \vec{\sigma}$ matrices are those of Minkowski space, $\vec{L} =
\vec{x} \times \vec{p}$,   $\vec{S} =  \vec{\sigma}/2 $  are the orbital and spin angular
momenta, respectively, and
\begin{equation}\label{sin10}
\nabla_\mu \Phi_{\rm G}  =   \frac{1}{2} \gamma_{\alpha \mu} (x) p^\alpha - \frac{1}{2}
\int_X^x dz^\lambda (\gamma_{\mu \lambda, \beta}(z) - \gamma_{\beta \lambda,
\mu}(z))p^\beta, \end{equation}
where $p^\mu$ is the momentum eigenvalue of the free particle. The term $
\vec{\omega}\cdot \vec{S}$ is the spin-rotation coupling term introduced by Mashhoon
\cite{mashh}.

\subsection{The Lense-Thirring effect for quantum systems}

An example of how a gravity induced phase is calculated can best be given by applying
(\ref{2b})-(\ref{2.4}), with $ \beta = 0$, to the LT effect \cite{caipap3}. This requires
knowledge of the particle paths and of the field $ \gamma_{\mu\nu}$.

Consider the physical situation illustrated in Fig.1.
\begin{figure}[top]
\begin{center}
\includegraphics[width=8cm,height=7.5cm]{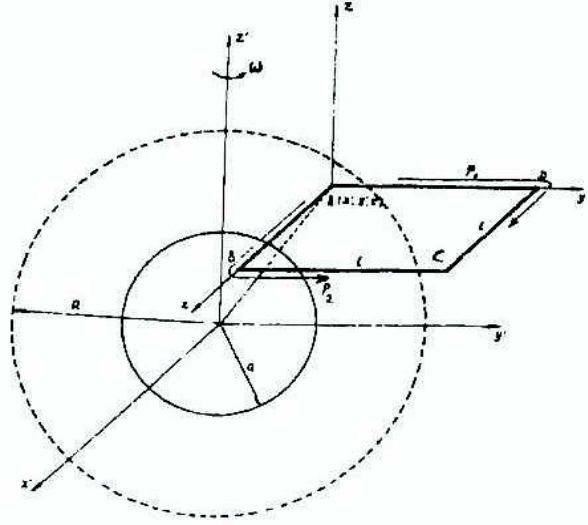}
\caption{Rotating, homogeneous, solid sphere and useful coordinates}
\label{fig:lenseth}
\end{center}
\end{figure}
A square interferometer of side $ l$ is represented by the path {\em ABCD} in the $
(xy)$-plane and a sphere of mass $M$ and radius $a$ is rotating about the $z'$-axis with
angular velocity $ \omega $. The spatial coordinates of the point $A$ at which a coherent
beam of particles is split are $(x', y', z')$ in the coordinate system $z^{\prime \mu}$.
For the sake of generality $A$ is taken a distance $R$ from the center of the sphere. The
beams interfere at $C$ after describing the paths $p_{1}\equiv ADC$ and $p_{2}\equiv
ABC$. Since the two coordinate systems $z^{\mu}$ and $z'^{ \mu}$ are at rest relative to
each other, one can choose $z^{0} = z^{\prime 0}$ and set the beam splitting time at $A$
to be $z^{0} = z^{\prime 0} =0$. It is sufficient to take $ \phi_{0}\propto
exp(ik_{\mu}x^{\mu})$, where $ k_{\mu}$ is the momentum of the particles of mass $ m$ in
the beams and $ k_{\mu}k^{\mu}= m^{2}$. The only non-vanishing values of $
\gamma_{\mu\nu}$ are \cite{lense}
\begin{eqnarray}\label{2SS}
\gamma_{00} &=&  \gamma_{ii}= -\frac{2M}{r}, \gamma_{01}= -\frac{4M\omega
a^{2}\left(y+y^{\prime}\right)}{5r^{3}}, \nonumber \\ \gamma_{02}&=&\frac{4M\omega
a^{2}\left(x+x^{\prime}\right)}{5r^{3}},
\end{eqnarray}
where $ r^{2} = (x+x^{\prime})^{2} +(y+y^{\prime})^{2} + (z+z^{\prime})^{2}$ and $R^{2}=
x^{\prime 2} + y^{\prime 2} +z^{\prime 2}$. The following expressions are also used below
\begin{eqnarray}\label{2AA}
\frac{1}{r}&=& \frac{1}{R} - \frac{x x^{\prime}}{R^{3}} - \frac{y y^{\prime}}{R^{3}} -
\frac{z z^{\prime}}{R^{3}} - \frac{1}{2 R^{3}} (x^{2} + y^{2} + z^{2}) + \nonumber \\ &&
\frac{3 x^{\prime 2} x^{2}}{2 R^{5}} + \frac{3 y^{\prime 2} y^{2}}{2 R^{5}} + \frac{3
z^{\prime 2} z^{2}}{2 R^{5}},\nonumber \\
 \frac{x^{\prime i}}{r^{3}}&=& \frac{x^{\prime i}}{R^{3}}
- \frac{3x^{\prime i} x^{\prime} x}{R^{5}} - \frac{3x^{\prime i} y^{\prime} y}{R^{5}} -
\frac{3x^{\prime i} z^{\prime} z}{R^{5}}, \nonumber \\  \frac{x^{\prime i} x^{\prime
j}}{r^{5}}&=& \frac{x^{\prime i} x^{\prime j}}{R^{5}}.
\end{eqnarray}

 The phase shift of the beams along the different arms of the
interferometer is given by
\begin{eqnarray}
\Delta \chi  &\equiv& \Delta \chi_{1} + \Delta \chi_{2}=
    \nonumber \\&&
 \frac{1}{4} \int^{x}_{A , p_{1}} dz^{\lambda} (\gamma_{\alpha \lambda , \beta}(z) -
\gamma_{\beta \lambda, \alpha}(z)) [(x^{\alpha} - z^{\alpha}) k^{\beta} - (x^{\beta} -
z^{\beta})k^{\alpha}] -\nonumber \\ & &  \frac{1}{4} \int^{x}_{A , p_{2}} dz^{\lambda}
(\gamma_{\alpha \lambda , \beta}(z) - \gamma_{\beta \lambda, \alpha}(z)) [(x^{\alpha} -
z^{\alpha}) k^{\beta} - (x^{\beta} - z^{\beta})k^{\alpha}] -\nonumber \\ & & \frac{1}{2}
\int^{x}_{A , p_{1}} dz^{\lambda} \gamma_{\alpha \lambda}(z) k^{\alpha} + \frac{1}{2}
\int^{x}_{A , p_{2}} dz^{\lambda} \gamma_{\alpha \lambda}(z) k^{\alpha}.
\end{eqnarray}
The calculation can be simplified by taking $ a =R$ and neglecting the contribution of
gravity to the motion of the particles in the beams. The latter choice is certainly
justified to first order in the WFA and for interferometers of laboratory dimensions.
Then path $p_{1}$ is described by
\begin{displaymath}
\begin{array}{lccc}
{\displaystyle 0 \leq z^{0} \leq \frac{\ell}{v} } &
  {\displaystyle   x= vz^{0} } &
y=0 \vspace{3mm} & {\displaystyle z=vz^{0} } \\
{\displaystyle \frac{\ell}{v} \leq z^{0} \leq \frac{2\ell}{v}} &
  {\displaystyle x= \ell } &
   {\displaystyle y= v z^{0} -\ell } &
   {\displaystyle z= 0 } \\

0 \leq x \leq \ell  & y = 0 &
   {\displaystyle z= 0 } \vspace{3mm} &
   {\displaystyle z^{0}= \frac{x}{v}}  \\
x= \ell  & 0 \leq y \leq \ell  &
    {\displaystyle z= 0} &
    {\displaystyle z^{0}= \frac{\ell}{v} + \frac{y}{v} }
\end{array}
\end{displaymath}
and $p_{2}$ by
\begin{displaymath}
\begin{array}{lccc}
{\displaystyle 0 \leq z^{0} \leq \frac{\ell}{v} } &
   x= 0 &  \vspace{3mm}
   {\displaystyle y=vz^{0}} &
   {\displaystyle z=0} \\
{\displaystyle \frac{\ell}{v} \leq z^{0} \leq \frac{2\ell}{v} } &
   {\displaystyle x= v z^{0} -\ell } &
   {\displaystyle y= \ell } &
   {\displaystyle z= 0} \\

{\displaystyle x=0 } & 0 \leq y \leq \ell &
   {\displaystyle z= 0 } &
   {\displaystyle z^{0}= \frac{y}{v} } \\
{\displaystyle 0 \leq x \leq \ell } & y= \ell &
   {\displaystyle z= 0 } &
   {\displaystyle z^{0}= \frac{\ell}{v} + \frac{x}{v } }.
\end{array}
\end{displaymath}
In addition for $p_{1}$ one has:
\begin{displaymath}
\begin{array}{lcc}
\mbox{at} \hspace*{2mm}  B: &
  x^{\mu}_{_{B}} = ( {\displaystyle \frac{\ell}{v}}, \ell , 0,0) \vspace{4mm}
   & k^{\mu}_{_{B}} = ( k^{0}, k, 0,0) \\
\mbox{at} \hspace*{2mm}  C: & x^{\mu}_{_{C}} = (
  {\displaystyle \frac{2\ell}{v}}, \ell ,
    \ell , 0) &
    k_{_{C}}^{\mu} = ( k^{0}, 0, k , 0)
\end{array}
\end{displaymath}
and for $p_{2}$
\begin{displaymath}
\begin{array}{lcc}
\mbox{at} \hspace*{2mm}  D: &
  x^{\mu}_{_{D}} = (
  {\displaystyle \frac{\ell}{v}}, 0, \ell , 0) \vspace{4mm}
  & k^{\mu}_{_{D}} = ( k^{0}, 0, k, 0) \\
\mbox{at} \hspace*{2mm}  C: & x^{\mu}_{_{C}} = (
  {\displaystyle \frac{2\ell}{v}}, \ell ,
    \ell, 0) &
    k_{_{C}}^{\mu} = ( k^{0}, k , 0, 0).
\end{array}
\end{displaymath}
Notice that the overall path described by the coherent beams is effectively closed in
space-time, as required by (\ref{2l}). On using the expressions for $ \gamma_{\mu\nu}$,
one finds
\begin{eqnarray}\label{pha}
\Delta \chi  &=&
 \frac{M \ell^{2}}{R^{3}} \frac{k^{0}}{v} \left(- x^{\prime } +
y^{\prime}+\frac{3x^{\prime 2} \ell }{2R^{2}} -\frac{3y^{\prime 2}\ell}{2R^{2}} \right) +
\nonumber \\ & &\frac{M \ell^{2}}{R^{3}} k \protect\left( - x^{\prime} + y^{\prime}+
 \frac{3x^{\prime 2}\ell}{2R^{2}}-
\frac{3y^{\prime 2}\ell}{2R^{2}}\protect\right) -\nonumber \\ & & \frac{2M  \ell^{2}
\omega a^{2}}{5R^{5}} \left( \frac{k}{v} + k^{0} \right) \left( 2R^{2}- 3 x^{\prime 2} -
3 y^{\prime 2} \right) {.}
\end{eqnarray}

 If the particles in the beam
have speed $v$, then in the non-relativistic approximation $ k^{0} \simeq m(1+
\frac{v^{2}}{2})$ and $ k \simeq m v $
  and $ \Delta\chi $
represents the phase measured by an observer co-moving with the interferometer relative
to which the sphere generating the LT field is spinning. The first term in $ \Delta\chi $
depends on $\omega$ and represents the LT effect experienced by the quantum particles. It
reaches its largest value when the interferometer is placed in the neighborhood of the
poles of the sphere $(x^{\prime}=y^{\prime}=0)$. The remaining terms represent
gravitational effects that are present even when $\omega=0$. These terms vanish when the
beam source is located at $x^{\prime}=y^{\prime}$ and, in particular at
$x^{\prime}=y^{\prime}=0$, at which positions the only contribution to the particle phase
shift is that of the LT field. For earth the first term can also be written, in normal
units, as
\begin{equation}
\Delta \chi_{ _{LT}} = \frac{2G}{c^{2}R^{3}_{\oplus}} J_{\oplus} \frac{m\ell}{\hbar}
[2R^{2}_{\oplus} - 3(x^{\prime 2} + y^{\prime 2})],   \end{equation} where $J_{\oplus}
=2M_{\oplus}R^{2}_{\oplus}\omega /5 $ is the angular momentum of earth (assumed spherical
and homogeneous) and $R_{\oplus}$ its radius. It is interesting to observe that $\Omega =
\frac{G}{2c^{2} R^{3}_{\oplus}} J_{\oplus}$ coincides with the effective LT precession
frequency of a gyroscope \cite{Weinberg,Will}. Since the precession frequency of a
gyroscope in orbit is $ \Omega=\frac{GJ_{\oplus}}{2c^2R_{\oplus}^3}$, one can also write
$ \Delta\chi=\Omega \Pi$, where $ \Pi=\frac{4ml^2}{\hbar}$ replaces the period of a
satellite in the classical calculation. Its value, $\Pi\sim 1.4\times10^{8}s $ for
neutron interferometers with $ l\sim10^2cm$, is rather high and yields
$\Delta\chi\sim10^{-7}rad$. This suggests that the development and use of large, heavy
particle interferometers would be particularly advantageous in attempts to measure the LT
effect.

\section{Inertial fields in particle accelerators}
\subsection{Spin-rotation coupling in g-2 experiments}
\setcounter{equation}{0}

 Prominent among the effects that can be derived from the covariant Dirac equation of
 Section 2.3 is the spin-rotation effect
described by Mashhoon \cite{mashh}. This effect is conceptually important. It extends our
knowledge of rotational inertia to the quantum level. It also yields different potentials
for different particles and for different spin states \cite{caipap2} and can not,
therefore, be considered universal.

The relevance of spin-rotation coupling to physical \cite{lloyd} and astrophysical
\cite{caipap2,papini94} processes has already been pointed out.

It is shown below that the spin-rotation effect plays an essential role in precise
measurements of the $g-2$ factor of the muon.

The experiment \cite{bailey,farley} involves muons in a storage ring consisting of a
vacuum tube, a few meters in diameter, in a uniform vertical magnetic field. Muons on
equilibrium orbits within a small fraction of the maximum momentum are almost completely
polarized with spin vectors pointing in the direction of motion. As the muons decay,
those electrons projected forward in the muon rest frame are detected around the ring.
Their angular distribution therefore reflects the precession of the muon spin along the
cyclotron orbits.

 The
calculations are performed in the rotating frame of the muon and do not therefore require
a relativistic treatment of inertial spin effects \cite{ryder}  . Then the vierbein
formalism yields (\ref{sin7}), or
\begin{equation}\label{gam}
  \Gamma_0=-\frac{1}{2}\,
  a_i\sigma^{0i}-\frac{1}{2}\,\omega_i\sigma^i\,,
\end{equation}
where
 \[
 \sigma^{0i}\equiv\frac{i}{2}\, [\gamma^0, \gamma^i]=i
 \left(\matrix{ \sigma^i & 0 \cr
                0 & -\sigma^i \cr }\right)\,
 \]
in the chiral representation of the usual Dirac matrices. The second term in (\ref{gam})
represents the Mashhoon effect. The first term drops out. The remaining contributions to
the Dirac Hamiltonian, to first order in $a_i$ and $\omega_i$, add up to
\cite{hehl,singh}
\begin{eqnarray}\label{hamiltonian-Ni}
  H &\approx & {\vec \alpha}\cdot {\vec p}+m\beta+\frac{1}{2}
  [({\vec a}\cdot {\vec x})({\vec p}\cdot {\vec \alpha})+
  ({\vec p}\cdot {\vec \alpha})({\vec a}\cdot {\vec x})] \\
  & & -{\vec \omega}\cdot \left({\vec L}+\frac{{\vec \sigma}}{2}\right)\,.
  \nonumber
\end{eqnarray}
For simplicity all quantities in $H$ are taken to be time-independent. They are referred
to a left-handed tern of axes rotating  about the $x_2$-axis in the clockwise direction
of motion of the muons. The $x_3$-axis is tangent to the orbits and in the direction of
the muon momentum. The magnetic field is $B_2=-B$. Only the Mashhoon term then couples
the helicity states of the muon. The remaining terms contribute to the overall energy $E$
of the states, and $H_0$ is the corresponding part of the Hamiltonian.

Before decay the muon states can be represented as
\begin{equation}\label{ps}
  |\psi(t)>=a(t)|\psi_+>+b(t)|\psi_->\,,
\end{equation}
where $|\psi_+>$ and $|\psi_->$ are the right and left helicity states of the Hamiltonian
$H_0$ and satisfy the equation
 \begin{equation}
 H_0|\psi_{+,-}>=E|\psi_{+,-}>\,.
 \end{equation}
The total effective Hamiltonian is $H_{eff}=H_0+H'$, where
\begin{equation}\label{3}
  H'=-\frac{1}{2}\,\omega_2\sigma^2+\mu B\sigma^2\,.
\end{equation}
$\displaystyle{\mu=\left(1+a_{\mu}\right)\mu_0}$ represents the total magnetic moment of
the muon and $\mu_0$ is the Bohr magneton. The effects of electric fields used to
stabilize the orbits and of stray radial electric fields can be cancelled by choosing an
appropriate muon momentum \cite{farley} and need not be considered.

The coefficients $a(t)$ and $b(t)$ in (\ref{ps}) evolve in time according to
\begin{equation}\label{4a}
  i\frac{\partial}{\partial t} \left(\matrix{ a(t) \cr
                b(t) \cr }\right)=M \left(\matrix{ a(t) \cr
                b(t) \cr }\right)\,,
\end{equation}
where $M$ is the matrix
\begin{equation}\label{5}
  M= \left[\matrix{ E-i\displaystyle{\frac{\Gamma}{2}} &
            \displaystyle{i\left(\frac{\omega_2}{2}-\mu B\right)}\cr
                \displaystyle{-i\left(\frac{\omega_2}{2}-\mu B\right)} &
                E-i\displaystyle{\frac{\Gamma}{2}} \cr }\right]
\end{equation}
and $\Gamma$ represents the width of the muon. The non-diagonal form of $M$ (when $B=0$)
implies that rotation does not couple universally to matter.

 $M$ has eigenvalues
\begin{eqnarray}
 h_1 &=& E-i\frac{\Gamma}{2}+\frac{\omega_2}{2}-\mu B \,, \nonumber \\
 h_2 &=& E-i\frac{\Gamma}{2}-\frac{\omega_2}{2}+\mu B \,,
\end{eqnarray}
and eigenstates
\begin{eqnarray}
 |\psi_1> &=&
 \frac{1}{\sqrt{2}}\,\left[i|\psi_+>+|\psi_->\right]\,, \nonumber
 \\
 |\psi_2> &=& \frac{1}{\sqrt{2}}\,\left[-i|\psi_+>+|\psi_->\right]
 \,.
\end{eqnarray}
The muon states that satisfy (\ref{ps}) and (\ref{4a}), and the condition $|\psi
(0)>=|\psi_->$ at $t=0$, are
\begin{eqnarray} \label{6a}
 |\psi(t)> &=& \frac{e^{-\Gamma t/2}}{2} e^{-iEt}
 \left\{
 i\left[e^{-i{\tilde \omega} t}
 -e^{i{\tilde \omega} t}\right]|\psi_+> \right.
 \\
 & & \left. + \left[e^{-i{\tilde \omega} t}
 +e^{i{\tilde \omega} t}\right]|\psi_-> \right\}
 \,, \nonumber
\end{eqnarray}
where
 \[
 {\tilde \omega}\equiv \frac{\omega_2}{2}-\mu B\,.
 \]
The spin-flip probability is therefore
\begin{eqnarray}\label{7}
  P_{\psi_-\to \psi_+}&=&|<\psi_+|\psi(t)>|^2 \\
     &=& \frac{e^{-\Gamma
  t}}{2}[1-\cos(2\mu B-\omega_2) t]\,. \nonumber
\end{eqnarray}
The $\Gamma$-term in (\ref{7}) accounts for the observed exponential decrease in electron
counts due to the loss of muons by radioactive decay \cite{farley}.

The spin-rotation contribution to $P_{\psi_-\to \psi_+}$ is represented by $\omega_2$
which is the cyclotron angular velocity $\displaystyle{\frac{eB}{m}}$ \cite{farley}. The
spin-flip angular frequency is then
 \begin{eqnarray}\label{omegafin}
 \Omega&=&2\mu B-\omega_2 \\
 &=&\left(1+\frac{g-2}{2}\right)\frac{eB}{m}-
 \frac{eB}{m} \nonumber \\
 &=& \frac{g-2}{2}\frac{eB}{m}= a_{\mu}\frac{eB}{m}\,,
 \end{eqnarray}
which is precisely the observed modulation frequency of the electron counts
\cite{picasso}. This result is independent of the value of the anomalous magnetic moment
of the particle. It is therefore the Mashhoon effect that evidences the $g-2$ term in
$\Omega$ by exactly cancelling, in $2\mu B$, the much larger contribution $\mu_0$ that
relates to fermions with no anomalous magnetic moment \cite{papini1}. The cancellation is
made possible by the non-diagonal form of $M$ and is therefore a direct consequence of
the violation of the equivalence principle.
 It is  significant that this effect is observed in an experiment that
has already provided crucial tests of quantum electrodynamics  and a test of Einstein's
time-dilation formula to better than a 0.1 percent accuracy. Recent versions of the
experiment \cite{carey,brown1,brown2} have improved the accuracy of the measurements from
$270 ppm$ to $1.3 ppm$ and  ultimately to $0.7 ppm$ \cite{brown3}. This, as well as
measurements of the Mashhoon effect using the Global Positioning System \cite{ashby},
bode well for studies involving spin, inertia and electromagnetic fields, or inertial
fields to higher order.

\subsection{Tests of parity and time reversal invariance}

The residual discrepancy $a_{\mu}(exp)-a_{\mu}(SM)=26\times 10^{-10}$ still existing
\cite{brown3} between the experimental and standard model values of the muon's  $a_{\mu}$
can be used to set an upper limit on $P$ and $T$ invariance violations in spin-rotation
coupling.

The possibility that discrete symmetries in gravitation be not conserved has been
considered by some authors \cite{schiff,leitner,dass,almeida}. Attention has in general
focused on the potential
\begin{equation}\label{1*}
U(\vec{r})=\frac{GM}{r}\left[\alpha_{1}\vec{\sigma}\cdot
\hat{r}+\alpha_{2}\vec{\sigma}\cdot \vec{v}+\alpha_{3}\hat{r}\cdot( \vec{v}\times
\vec{\sigma})\right],
\end{equation}
which applies to a particle of generic spin $\vec{\sigma}$. The first term, introduced by
Leitner and Okubo \cite{leitner}, violates the conservation of $P$ and $T$. The same
authors determined the upper limit $\alpha_{1}\leq 10^{-11}$ from the hyperfine splitting
of the ground state of hydrogen. The upper limit $\alpha_{2}\leq 10^{-3}$ was determined
in Ref.\cite{almeida} from SN 1987A data. The corresponding potential violates the
conservation of $P$ and $C$. Conservation of $C$ and $T$ is violated by the last term,
while (\ref{1*}), as a whole, conserves $CPT$. There is, as yet, no upper limit on
$\alpha_{3}$. These studies can be extended to the Mashhoon term.

Assume, in fact, that the coupling of rotation to $\mid\psi_{+}>$ differs in strength
from that to $\mid\psi_{-}>$ \cite{pap02}. Then the Mashhoon term can be altered by means
of a matrix $A=\left(\matrix{\kappa_{1}&0\cr 0&\kappa_{2}\cr}\right)$ that reflects the
different coupling of rotation to the two helicity states. The total effective
Hamiltonian is $H_{eff}=H_{0}+H'$, where
\begin{equation}\label{4*}
H'=-\frac{1}{2}A \omega_{2}\sigma_{2}+\mu B\sigma_{2}.
\end{equation}
A violation of $P$ and $T$ in (\ref{4*}) would arise through $\kappa_{2}-\kappa_{1}\neq
0$. The constants $\kappa_{1}$ and $\kappa_{2}$ are assumed to differ from unity by small
amounts $\epsilon_{1}$ and $\epsilon_{2}$.

The muon states before decay are again as in (\ref{ps}) and  the coefficients $a(t)$ and
$b(t)$  evolve in time according to (\ref{4a}), but now the matrix $ M$ is replaced by
\begin{equation}\label{6*}
\tilde{M}=\left(\matrix{E-i\frac{\Gamma}{2}& i\left(\kappa_{1}\frac{\omega_{2}}{2}-\mu
B\right)\cr -i\left(\kappa_{2}\frac{\omega_{2}}{2}-\mu B\right)&
E-i\frac{\Gamma}{2}\cr}\right).
\end{equation}
The spin-rotation term, that is off-diagonal in (\ref{6*}), violates Hermiticity and $T$,
$P$ and $PT$, as shown in \cite{pap02} and, in a general way, in \cite{scol}, while
nothing can be said about $CPT$ conservation which requires $H_{eff}$ to be Hermitian
\cite{kennysachs,sachs}. Because of the non-Hermitian nature of (\ref{4*}), one expects
$\Gamma$ itself to be non-Hermitian. The resulting corrections to the width of the muon
are, however, of second order in the $\epsilon$'s and are neglected.

$\tilde{M}$ has eigenvalues
\begin{eqnarray}
h_{1}&=&E-i\frac{\Gamma}{2}+R \nonumber \\ h_{2}&=&E-i\frac{\Gamma}{2}-R,
\end{eqnarray}
where
\begin{equation}
R=\sqrt{\left(\kappa_{1}\frac{\omega_{2}}{2}-\mu
B\right)\left(\kappa_{2}\frac{\omega_{2}}{2}-\mu B\right)},
\end{equation}
and eigenstates
\begin{eqnarray}
|\psi_{1}>&=&b_{1}\left[\eta_{1}|\psi_{+}>+|\psi_{-}>\right],\nonumber \\
|\psi_{2}>&=&b_{2}\left[\eta_{2}|\psi_{+}>+|\psi_{-}>\right].
\end{eqnarray}
One also finds
\begin{eqnarray}
|b_{1}|^{2}&=&\frac{1}{1+|\eta_{1}|^{2}}\nonumber \\
|b_{2}|^{2}&=&\frac{1}{1+|\eta_{2}|^{2}}
\end{eqnarray}
and
\begin{equation}
\eta_{1}=-\eta_{2}=\frac{i}{R}\left(\kappa_{1}\frac{\omega_{2}}{2}-\mu B\right).
\end{equation}
Then the muon states (\ref{ps}) are
\begin{eqnarray}
|\psi(t)>&=&\frac{1}{2}e^{-iEt-\frac{\Gamma t}{2}}[-2i\eta_{1} \sin Rt |\psi_{+}>+
\nonumber \\
         & &2\cos Rt
         |\psi_{-}>],
\end{eqnarray}
where the condition $|\psi(0)>=|\psi_{-}>$ has been applied. The spin-flip probability is
therefore
\begin{eqnarray}\label{D*}
P_{\psi_{-}\rightarrow \psi_{+}}&=&|<\psi_{+}|\psi(t)>|^{2}\nonumber
\\&=&\frac{e^{-\Gamma t}}{2}\frac{\kappa_{1}\omega_{2}-2\mu B}{\kappa_{2}\omega_{2}-2\mu
B}\left[1-\cos 2Rt\right].
\end{eqnarray}
This equation and $ \kappa_{1}=\kappa_{2}=1$, yield (\ref{6a}) and (\ref{7}) that provide
the appropriate description of the spin-rotation contribution to the spin-flip transition
probability. Notice that the case $\kappa_{1}=\kappa_{2}=0$ (vanishing spin-rotation
coupling) gives
\begin{equation}\label{AB}
P_{\psi_{-}\rightarrow \psi_{+}}=\frac{e^{-\Gamma
t}}{2}\left[1-\cos(1+a_{\mu})\frac{eB}{m}\right]
\end{equation}
and does not therefore agree with the results of the $g-2$ experiments. Hence the
necessity of accounting for spin-rotation coupling whose contribution cancels the factor
$\frac{eB}{m}$ in (\ref{AB})\cite{papini1}.

 Substituting $\kappa_{1}=1+\epsilon_{1},
\kappa_{2}=1+\epsilon_{2}$ into (\ref{D*}), one finds
\begin{equation}\label{11}
P_{\psi_{-}\rightarrow\psi_{+}}\simeq\frac{e^{-\Gamma t}}{2}[1-\cos
\frac{eB}{m}(a_{\mu}-\epsilon)t] ,
\end{equation}
where $\epsilon=\frac{1}{2}(\epsilon_{1}+\epsilon_{2})$. One may attribute the
discrepancy between $a_{\mu}(exp)$ and $a_{\mu}(SM)$ to a violation of the conservation
of the discrete symmetries by the spin-rotation coupling term in (\ref{4*}). The upper
limit on the violation of $P,T$ and $PT$ is derived from (\ref{11}) assuming that the
deviation from the current value of $a_{\mu}(SM)$ is wholly due to $\epsilon$, and
therefore is $26\times 10^{-10}$.

\section{Maximal acceleration}
\setcounter{equation}{0} In the 1980's, Caianiello and collaborators \cite{ma} developed
a geometrical model of quantum mechanics in which quantization is interpreted as
curvature of the eight-dimensional space-time tangent bundle $TM = M_{4}\otimes TM_{4}$,
where $M_4$ is the usual flat space--time manifold, of metric $\eta_{\mu\nu}$. In this
space the standard operators of the Heisenberg algebra are represented as covariant
derivatives and the quantum commutation relations are interpreted
 as components of the curvature tensor. The usual Minkowski line element is
 replaced in the model by
the infinitesimal element of distance in the eight-dimensional space-time tangent bundle
$TM$
\begin{equation}\label{4.1}
d\tau^2=\eta_{AB}dX^AdX^B \qquad\qquad A, \, B = 0, \ldots, 7,
\end{equation}
where, in normal units, $ \eta_{AB}=\eta_{\mu\nu}\otimes \eta_{\mu\nu}\,{,}
X^{A}=\left(x^{\mu},\frac{c^2}{{\cal A}_m} \frac{dx^{\mu}}{ds}\right), \mu=0,\ldots,3 ,
x^{\mu}=(ct,\vec{x}), dx^{\mu}/ {ds}=\dot x^{\mu}$ is the relativistic four-velocity and
${\cal A}_m $ is a constant. In the model the symmetry between configuration and momentum
space representations of field theory (Born reciprocity theorem) is automatically
satisfied. The invariant line element (\ref{4.1}) can be written in the form $$
d\tau^2=\eta_{\mu\nu}dx^{\mu}dx^\nu +{1\over {\cal A}_m^2} \eta_{\mu\nu}d\dot
x^{\mu}d\dot x^\nu= $$
\begin{equation}\label{4.2}
= \left[1+\frac{\ddot x_\mu\ddot x^\mu}{{\cal A}_m^2}\right] ds^{2}\equiv \sigma^{2}(x)
ds^{2} ,
\end{equation}
where all proper accelerations are normalized to ${\cal A}_m$, referred to as maximal
acceleration, very much like velocities are normalized to their upper value $ c$. Though
${\cal A}_m$ is, a priori, arbitrary, a value for it can be derived from quantum
mechanics \cite{ca}. With some modifications and additions \cite{pw,pap03}, Caianiello's
argument can be re-stated as follows.

If two observables $ \hat{f}$ and $ \hat{g}$ obey the commutation relation
\begin{equation}\label{al}
\left[\hat{f},\hat{g}\right]= - i \hbar \hat{\alpha},
\end{equation}
where $ \hat{\alpha}$ is a Hermitian operator, then their uncertainties
\begin{eqnarray}
 \left(\Delta f\right)^{2}&=&
<\Phi\mid\left(\hat{f}-<\hat{f}>\right)^{2}\mid\Phi>\\ \nonumber
 \left(\Delta g\right)^{2}&=&
<\Phi\mid\left(\hat{g}-<\hat{g}>\right)^{2}\mid\Phi>
\end{eqnarray}
 also satisfy the inequality
\begin{equation}
\left(\Delta f\right)^{2}\cdot\left(\Delta g\right)^{2}\geq
\frac{\hbar^{2}}{4}<\Phi\mid\hat{\alpha}\mid\Phi>^{2},
\end{equation}
or
\begin{equation}\label{be}
\Delta f\cdot\Delta g\geq \frac{\hbar}{2} \mid<\Phi\mid\hat{\alpha}\mid\Phi>\mid .
\end{equation}
Using Dirac's analogy between the classical Poisson bracket $ \left\{f,g\right\}$ and the
quantum commutator \cite{land}
\begin{equation}
\left\{f,g\right\} \rightarrow\frac{1}{i\hbar}\left[\hat{f},\hat{g}\right],
\end{equation}
one can take $ \hat{\alpha}=\left\{f,g\right\}\hat{\mathbf{1}}$. With this substitution,
Eq.(\ref{al}) yields the usual momentum-position commutation relations. If in particular
$ \hat{f}=\hat{H}$, then Eq.(\ref{al}) becomes
\begin{equation}
\left[\hat{H},\hat{g}\right]= - i \hbar \left\{H,g\right\}\hat{\mathbf{1}},
\end{equation}
Eq.(\ref{be}) gives \cite{land}
\begin{equation}\label{&}
\Delta E\cdot \Delta g \geq \frac{\hbar}{2} \mid \left\{H,g\right\}\mid
\end{equation}
and
\begin{equation}\label{&&}
\Delta E\cdot\Delta g\geq \frac{\hbar}{2}\mid\frac{dg}{dt}\mid ,
\end{equation}
when $ \frac{\partial g}{\partial t}=0 $. Eqs.(\ref{&}) is  Ehrenfest theorem. Criteria
for its validity are discussed at length in the literature \cite{mess,land}.
Eq.(\ref{&&}) implies that $\Delta E =0$ when the quantum state of the system is an
eigenstate of $\hat{H}$. In this case $\frac{dg}{dt}= 0$.

If $g \equiv v(t)$ is the (differentiable) velocity expectation value of a particle whose
average energy is $ E$, then Eq.(\ref{&&}) gives
\begin{equation}\label{4.7}
\mid \frac{dv}{dt}\mid \leq \frac{2}{\hbar} \Delta E \cdot \Delta v(t) .
\end{equation}
 In general \cite{sha}
\begin{equation}
 \Delta v =
\left(<v^{2}>-<v>^{2}\right)^{\frac{1}{2}}\leq v_{max}\leq c .
\end{equation}
 Caianiello's additional assumption, $ \Delta E \leq E$, has so far remained unjustified.
In fact, Heisenberg's uncertainty relation
\begin{equation}\label{yy}
\Delta E \cdot \Delta t\geq \hbar/2 ,
\end{equation}
 that follows from (\ref{4.7}) by writing $ \Delta t =
\Delta v /|dv/dt| $, seems to imply that, given a fixed energy E, a state can be
constructed with arbitrarily large $\Delta E$, contrary to Caianiello'assumption. An
upper bound on $\Delta E$ can be found, however, if $E$ is taken to represent the fixed
\textit{average} energy measured from an origin $ E_{min}$. In what follows $E_{min}=0$
for simplicity. Then the correct interpretation of (\ref{yy}) is that a quantum state
with spread in energy $ \Delta E$ takes a time $ \Delta t\geq \frac{\hbar}{2 \Delta E}$
to evolve to a distinguishable (orthogonal) state. This evolution time must satisfy the
more stringent limit \cite{marg}
\begin{equation} \label{&&&}
\Delta t\geq \frac{\hbar}{2E} ,
\end{equation}
which determines a maximum speed of orthogonality evolution  \cite{beck}. Obviously, both
limits (\ref{yy}) and (\ref{&&&}) can be achieved only for $ \Delta E = E$, while spreads
$ \Delta E > E $, that would make $ \Delta t $ smaller, are precluded by (\ref{&&&}).
This effectively
 restricts $\Delta E$ to values
$\Delta E \leq E$, as conjectured by Caianiello. One can now derive an upper limit on the
value of the proper acceleration. In fact, in the instantaneous rest frame of the
particle, where the acceleration is largest \cite{pw}, $ E = mc^{2}$ and (\ref{4.7})
gives
\begin{equation}\label{8}
\mid \frac{dv}{dt}\mid \leq 2\frac{mc^{3}}{\hbar}\equiv {\mathcal A}_{m} .
\end{equation}

 It also follows that in the rest frame of
the particle, where $ \frac{d^{2}x^{0}}{ds^{2}}= 0 $, the absolute value of the proper
acceleration is \cite{pw,steph}
\begin{equation}\label{9}
\left(\mid
\frac{d^{2}x^{\mu}}{ds^{2}}\frac{d^{2}x_{\mu}}{ds^{2}}\mid\right)^{\frac{1}{2}}=
\left(\mid\frac{1}{c^{4}}\frac{d^{2}x^{i}}{dt^{2}}\mid \right)^{\frac{1}{2}}\leq
\frac{\mathcal{A}_{m}}{c^{2}} .
\end{equation}
Eq.(\ref{9}) is a Lorentz invariant. The validity of (\ref{9}) under Lorentz
transformations is therefore assured.

Result (\ref{&&&}) can also be used to extend (\ref{8}) to include the average length of
the acceleration $ <a>$. If, in fact, $ v(t)$ is differentiable, then fluctuations about
its mean are given by
\begin{equation}\label{del}
\Delta v \equiv v-<v>\simeq \left(\frac{dv}{dt}\right)_{0}\Delta
t+\left(\frac{d^{2}v}{dt^{2}}\right)_{0}\left(\Delta t\right)^{2}+...  .
\end{equation}
Eq.(\ref{del}) reduces to $ \Delta v\simeq \mid\frac{dv}{dt}\mid \Delta t =<a> \Delta t$
for sufficiently small values of $\Delta t$, or when $\mid\frac{dv}{dt}\mid$ remains
constant over $\Delta t$. Eq.(\ref{&&&}) then yields
\begin{equation}\label{ga}
<a> \leq \frac{2cE}{\hbar}
\end{equation}
and again (\ref{8}) follows \cite{pap03}.

Classical and quantum arguments supporting the existence of a maximal acceleration have
long been discussed in the literature \cite{prove}. MA also appears in the context of
Weyl space \cite{pap} and of a geometrical analogue of Vigier's stochastic theory
\cite{jv}.

MA has been used to obtain model independent limits on the mass of the Higgs boson
\cite{Kuwata} and on the stability of white dwarfs and neutron stars \cite{shadows}.

It is significant that a limit on the acceleration also occurs in string theory. Here the
upper limit manifests itself through Jeans-like instabilities \cite{gsv} which occur when
the acceleration induced by the background gravitational field is larger than a critical
value $a_c = (m\alpha)^{-1}$for which the string extremities become causally disconnected
\cite{gasp}. $m$ is the string mass and $\alpha$ is the string tension. Frolov and
Sanchez \cite{fs} have then found that a universal critical acceleration $a_c =
(m\alpha)^{-1}$ must be a general property of strings.

Recently Castro \cite{castro} has derived the same MA limit (\ref{8}) from Clifford
algebras in phase space and Schuller \cite{schul} has rigorously shown that special
relativity has a MA extension.

Applications of the Caianiello model range from cosmology to particle physics. A sample
of pertinent references can be found in \cite{caianie}. Clearly (\ref{4.2}) implies that
the effective space-time metric experienced by accelerated particles is $
\tilde{g}_{\mu\nu}=\sigma^{2}\eta_{\mu\nu}$ and is therefore altered by MA corrections
that induce curvature, violate the equivalence principle and make the metric observer
dependent as conjectured by Gibbons and Hawking \cite{GIB}. These corrections vanish in
the classical limit $({\cal A}_m)^{-1}=\hbar /(2mc^{3})\rightarrow 0 $, as expected.

Recent advances in high resolution spectroscopy are now allowing Lamb shift mesearements
of unprecedented precision, leading in the case of simple atoms and ions to the most
stringent tests of quantum electrodynamics (QED). MA corrections due to the metric
(\ref{4.2}) appear directly in the Dirac equation for the electron that must now be
written in covariant form and referred to a local Minkowski frame by means of the
vierbein field $e_{\mu}^{\,\,\,\, a}(x)$. From (\ref{4.2}) one finds $e_{\mu}^{\,\,\,\,
a} =\sigma(x)\delta_{\mu}^{\,\,\,\, a}$, where Latin indices refer to the locally
inertial frame and Greek indices to a generic non-inertial frame. The covariant matrices
$\gamma^{\mu}(x)$ satisfy the anticommutation relations ~$\{\gamma^{\mu}(x),
\gamma^{\nu}(x)\}$ $=2\tilde g^{\mu\nu}(x)$, while the  covariant derivative ${\cal
D}_{\mu}\equiv \partial_{\mu}+\omega_{\mu}$ contains the total connection $\omega_{\mu}=
\frac{1}{2}\sigma^{ab}\omega_{\mu ab}$, where
$\sigma^{ab}=\frac{1}{4}\,[\gamma^a,\gamma^b]$,
$\omega_{\mu\,\,\,\,b}^{\,\,\,\,a}=(\Gamma_{\mu\nu}^{\lambda}\,
e_{\lambda}^{\,\,\,\,a}-\partial_{\mu}e_{\nu}^{\,\,\,\,a}) e^{\nu}_{\,\,\,\,b}$ and
$\Gamma_{\mu\nu}^{\lambda}$ represent the usual Christoffel symbols. For conformally flat
metrics $\omega_{\mu}$ takes the form $\omega_{\mu}=\frac{1}{
\sigma}\sigma^{ab}\eta_{a\mu}\sigma_{,b}$. By using the transformations
$\gamma^{\mu}(x)=e^{\mu}_{\,\,\,\,a}(x)\gamma^a$ so that $\gamma^{\mu}(x)=\sigma^{-1}
(x)\gamma^{\mu}$, where $\gamma ^{\mu}$ are the usual constant Dirac matrices, the Dirac
equation can be written in the form
\begin{equation}\label{eq3}
\left[ i\hbar\gamma^{\mu}\left(\partial_{\mu}+i\frac{e}{\hbar c}A_{\mu}\right)
+i\frac{3\hbar}{2}\gamma^{\mu}(\ln\sigma)_{,\mu} -mc\sigma (x)\right]\psi(x)=0\,{.}
\end{equation}
From (\ref{eq3}) one obtains the Hamiltonian
\begin{equation}\label{eq4}
H= - i\hbar c\vec{\alpha}\cdot \vec{\nabla} + e \gamma^0 \gamma^{\mu}A_{\mu}(x) -
i\frac{3\hbar c}{2} \gamma^0 \gamma^{\mu}(\ln\sigma )_{,\mu} + mc^2\sigma(x)\gamma^0\,{,}
\end{equation}
which is in general non--Hermitian \cite{PAR}. However, when one splits the Dirac spinor
into large and small components, the only non-Hermitian term is $(\ln\sigma )_{,0}$. If
$\sigma$ varies slowly in time, or is time-independent, as in the present case, this term
can be neglected and Hermiticity is recovered. Here the nucleus is considered to be
point-like and its recoil is neglected.

In QED the Lamb shift corrections are usually calculated by means of a non--relativistic
approximation \cite{ITZ} which is also followed here \cite{lamb1,lamb2}. For the electric
field $E(r)=kZe/r^2 (k=1/4\pi\epsilon_{0})$, the conformal factor becomes $
\sigma(r)=(1-\left(\frac{r_0}{r}\right)^4)^{1/2}$, where $ r_0\equiv (kZe^2/m{\cal
A}_m)^{1/2}\sim \sqrt{Z}\,2.3\cdot 10^{-14}\mbox{m} $ and $r>r_0$. The calculation of
$\ddot{x}^{\mu}$ is performed classically in a non--relativistic approximation. This is
justified because for the electron $v/c$ is at most $\sim 10^{-3}$. Neglecting
contributions of the order $O({\cal A}_m^{-4})$, $ \sigma (r)\sim 1-(1/2)(r_0/r)^4 $.
This expansion requires that in the following only those values of $r$ be chosen that are
above a cut--off $\Lambda$, such that for $r>\Lambda >r_0$ the validity of the expansion
is preserved. The actual value of $\Lambda$ is chosen below. The length $r_0$ has no
fundamental significance in QED and depends in general on the details of the acceleration
mechanism. It is only the distance at which the electron would attain, classically, the
acceleration ${\cal A}_m$ irrespective of the probability of getting there.

By using the expansion for $\sigma (r)$ in (\ref{eq4}) one finds that all MA effects are
contained in the perturbation terms
\begin{equation}\label{eq8}
H_{r_0}=-\frac{mc^2}{2}\left(\frac{r_0}{r}\right)^4\beta+ i\frac{3\hbar
c}{4}r_0^4\vec{\alpha}\cdot\vec{\nabla} \frac{1}{r^4}\equiv {\cal H}+{\cal
H}^{\prime}\,{.}
\end{equation}
By splitting $\psi(x)$ into large and small components $\varphi$ and $\chi$ and using
$\chi=-i(\hbar/2mc)\vec{\sigma}\cdot\vec{\nabla}\varphi\ll\varphi$ one obtains for the
perturbation due to ${\cal H}$
\begin{equation}\label{eq9}
\delta {\cal E}_{nlm}\simeq -\frac{mc^2}{2}r_0^4\int d^3\vec{r}\frac{1}{r^4}
\varphi^{*}_{nlm}\varphi_{nlm}\,{.}
\end{equation}
The perturbation due to ${\cal H}^{\prime}$ vanishes. In (\ref{eq9}) $\varphi_{nlm}$ are
the well known eigenfunctions for one--electron atoms. The integrations over the angular
variables in (\ref{eq9}) can be performed immediately and yield
\begin{eqnarray}
\delta {\cal E}_{20} &=& -\frac{mc^2}{16}\left(\frac{r_0}{a_0}\right)^4
     \left\{\left[4\left(\frac{a_0}{\Lambda}\right)+1\right]e^{-\Lambda/a_0}
-8E_1\left(\frac{\Lambda}{a_0}\right)\right\}\,{,} \label{eq12}\\ \delta {\cal
E}_{21}&=&-\frac{mc^2}{48}\left(\frac{r_0}{a_0}\right)^4
 e^{-\Lambda/a_0} \,{,}  \label{eq13} \\
\delta{\cal E}_{10}&=&-2mc^2\left(\frac{r_0}{a_0}\right)^4\left[
\left(\frac{a_0}{\Lambda}\right)e^{-2\Lambda/a_0}-
            2E_1\left(\frac{2\Lambda}{a_0}\right)\right]\,{,} \label{eq14}
\end{eqnarray}
where $ E_1 (x)=\int_{1}^{\infty} dy\,e^{-xy}/y$ and $a_0$ is the Bohr radius divided by
$Z$. In order to calculate the $2S-2P$ Lamb shift corrections it is now necessary to
choose the value of the cut--off $\Lambda$. While in QED Lamb shift and fine structure
effects are cut--off independent, the values of the corresponding MA corrections increase
when $\Lambda$ decreases. This can be understood intuitively because the electron finds
itself in regions of higher electric field at smaller values of $r$. $\Lambda$ is a
characteristic length of the system. It must also represent a distance from the nucleus
that can be reached by the electron whose acceleration and relative perturbations depend
on the position attained. One may tentatively choose $\Lambda\sim a_0$. According to the
wave functions involved, the probability that the electron be at this distance ranges
between $0.1$ and $0.5$. Smaller values of $\Lambda$ lead to larger acceleration
corrections, but are reached with much lower probabilities. This is the case of the
Compton wavelength of the electron whose use as a cut--off is therefore ruled out in the
present context. For $\Lambda\sim a_0$, Eqs. (\ref{eq12})-(\ref{eq14}) give the
corrections to the levels $2S, 2P$ and $1S$ ($Z=1$) $\delta {\cal E}_{20}\sim
-22.96\,\mbox{kHz}$, $\delta {\cal E}_{21}\sim -33.42\,\mbox{kHz}$, $\delta {\cal
E}_{10}\sim -325.45\, \mbox{kHz}$, yielding the Lamb shift correction $\delta {\cal
E}_L=\delta {\cal E}_{20}-\delta {\cal E}_{21} \sim +10.46\,\mbox{kHz}$. A fully
relativistic calculation \cite{chen} gives $ \delta {\cal E}_{L}\sim 11.37 kHz$. The MA
corrections are comparable in magnitude with those of QED at order $ \alpha^{7}$, where $
\alpha $ is the fine structure constant. The agreement between MA corrections and
experiment \cite{RP1,RP2} is at present very good \cite{lamb2} for the $2S-2P$ Lamb shift
in hydrogen ($\sim 7\mbox{kHz}$) and comparable with the agreement of experiments with
standard QED with and without two-loop corrections \cite{PAC}. The agreement is also good
for the $\frac{1}{4}L_{1S}-\frac{5}{4}L_{2S}+L_{4S}$ Lamb shift in hydrogen and
comparable, in some instances, with that between experiment and QED ($\sim 30
\mbox{kHz}$) \cite{lamb2,EXA}. Finally, the MA corrections \cite{lamb2} improve the
agreement between experiment \cite{AVV} and theory by $\sim 50\%$ for the $2S-2P$ shift
in $He^+$.

\section{Conclusions}

Inertia and gravity induced quantum phases, helicity oscillations of particles in
accelerators and storage rings and MA corrections in quantum processes are all effects
that may occur well before the onset of quantum gravity. They represent research areas
where both theoretical and experimental developments are possible.

The sensitivity of measurements in g-2 and Lamb shift experiments can respectively set
upper limits on violations of P and T invariance in spin-rotation coupling and on the
magnitude of MA corrections.

Further advances in these fields as well as in heavy particle interferometry, would
greatly help in filling a gap of over forty orders of magnitude between planetary scales,
over which Einstein's views on inertia and gravity are tested, and Planck length.

\end{document}